# Hydrothermal formation of Clay-Carbonate alteration assemblages in the Nili Fossae region of Mars

Adrian J. Brown[*1], Simon J. Hook[2], Alice M. Baldridge[2], James K. Crowley[3], Nathan T. Bridges[2], Bradley J. Thomson[4], Giles M. Marion[5], Carlos R. de Souza Filho[6], Janice L. Bishop[1],

[1] SETI Institute, 515 N. Whisman Rd, Mountain View, CA 94043, USA
[2] Jet Propulsion Laboratory, 4800 Oak Grove Dr, CA 91109, USA
[3] P.O. Box 344, Lovettsville, VA 20180, USA
[4] Johns Hopkins University Applied Physics Laboratory, Laurel, MD, 20723, USA
[5] Desert Research Institute, 2215 Raggio Pkwy, Reno, NV 89512, USA
[6] Universidade Estadual de Campinas, Campinas, São Paulo, Brasil

Corresponding author:
Adrian Brown
SETI Institute
515 N. Whisman Rd Mountain View, CA 94043
ph. 650 810 0223
fax. 650 968 5830
email. abrown@seti.org

## ABSTRACT

The Compact Reconnaissance Imaging Spectrometer for Mars (CRISM) has returned observations of the Nili Fossae region indicating the presence of Mg-carbonate in small (<10km sq$^2$), relatively bright rock units that are commonly fractured (Ehlmann et al., 2008b). We have analyzed spectra from CRISM images and used co-located HiRISE images in order to further characterize these carbonate-bearing units. We applied absorption band mapping techniques to investigate a range of possible phyllosilicate and carbonate minerals that could be present in the Nili Fossae region. We also describe a clay-carbonate hydrothermal alteration mineral assemblage in the Archean Warrawoona Group of Western Australia that is a potential Earth analog to the Nili Fossae carbonate-bearing rock units. We discuss the geological and biological implications for hydrothermal processes on Noachian Mars.

## KEYWORDS

---

[*] corresponding author, email: abrown@seti.org

# INTRODUCTION

Data from the CRISM instrument has recently been used to detect the presence of Mg-carbonate minerals in the Nili Fossae region of Mars (Ehlmann et al., 2008b). Nili Fossae displays a diverse range of minerals in a classic 'low dust cover' volcanic province that contains exposed rock units dating back to the Noachian period of Mars (Mustard et al., 1993; Ehlmann et al., 2007; Mangold et al., 2007; Mustard et al., 2007; Ehlmann et al., 2009). Ehlmann et al. (2008b) proffered four potential formation scenarios for the carbonate-bearing unit – 1.) groundwater percolating through fractures altering olivine to Mg-carbonate at slightly elevated temperatures, 2.) olivine-rich material, heated by impact or volcanic processes, was deposited on top of a water-bearing phyllosilicate rich unit and initiated hydrothermal alteration along the contact, 3.) olivine-rich rocks were weathered to carbonate at surface (cold) temperatures in a manner similar to olivine weathering of meteorites in Antarctica, and 4.) the carbonate precipitated from shallow ephemeral lakes.

Clay-carbonate alteration formation hypothesis. Here we further develop and slightly modify Ehlmann's hydrothermal emplacement hypothesis (#2). We propose that the phyllosilicate (possibly talc) and overlying carbonate bearing unit was formed at the same time by a single hydrothermal event (Brown et al., 2008a). In this scenario, the phyllosilicate (argillic) and carbonate (propylitic) zones reflect different temperature zones achieved during hydrothermal alteration.

The Nili Fossae region contains large amounts of olivine (Hoefen et al., 2003; Mustard et al., 2005) apparently in volcanic basalt (Hamilton and Christensen, 2005). In order to form the observed Mg-carbonate unit and stratigraphically lower Mg-phyllosilicate (talc)-bearing unit, we propose that the carbonate-bearing units have undergone alteration in a circum-neutral (6-8) pH environment, similar to regions of the mafic/ultramafic Warrawoona Group (Figure 1) in Western Australia (Brauhart et al., 2001; Van Kranendonk et al., 2002; Brown et al., 2005).

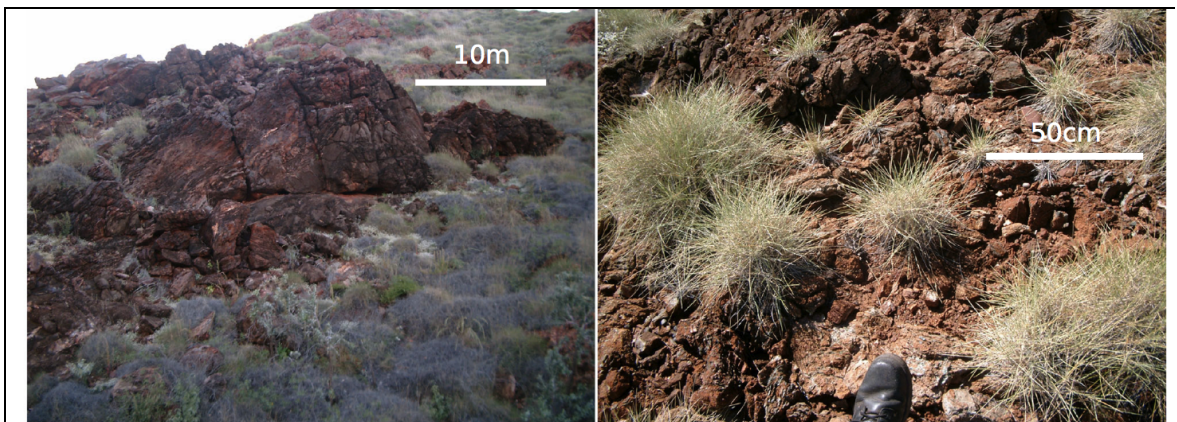

Figure 1 – Clay-carbonate altered basalt outcrop in the Pilbara region of Western Australia (a) from a distance and (b) close up. Foot for scale.

# BACKGROUND

## Geological Setting of the North Pole Dome Region of the Pilbara

In October 2002 an airborne hyperspectral imaging dataset was acquired over the North Pole Dome Region of the East Pilbara Granite-Greenstone terrain, where significant outcrops of the mafic/ultramafic Warrawoona Group are located (Brown et al., 2005). The North Pole Dome region provides excellent exposure of low-grade (prehnite-pumpellyite to greenschist) metamorphism. In the Warrawoona Group, there are large areas where weathering rinds do not obscure inherent mineralogy from airborne reflectance spectrometers, making it an excellent study region for proving remote sensing technologies.

The rocks of the Warrawoona Group constitute two komatiitic-thoelitic-felsic-chert volcanic successions which have ages spanning 3.515-3.426Ga - as each succession gets younger in age, it gets progressively less mafic (Van Kranendonk et al., 2002). The hyperspectral VNIR signature of talc has been used to map a komatiite layer around the North Pole Dome in the Apex Basalt member of the Warrawoona Group (Brown et al., 2004b). The Apex Basalt overlies the stromatolite-bearing 3.49 Ga Dresser Formation chert-barite unit, which probably represents the late stage of an active volcanic caldera (Van Kranendonk et al., 2008). The komatiitic 3.46 Ga Apex Basalt probably represents resumption of distal volcanic activity following a ~20k year hiatus. Talc-carbonate Hydrothermal alteration of the Apex Basalt was either achieved on emplacement of the komatiite or when the overlying theolitic 3.46Ga Mt. Ada Basalt unit was emplaced.

Komatiite lavas form when high-temperature (~1400-1600°C), low viscosity (0.1-1Pa), mantle derived, ultramafic lavas are extruded and flow turbulently at the surface. Komatiites are found almost exclusively in Archean shield areas due to the higher heat of the Earth's mantle during that period (Campbell et al., 1989). Komatiite rocks on Earth have high Mg contents (> 9% MgO by weight – those in the Warrawoona Group have up to 29% MgO), which may be a significant factor on Mars since only Mg-carbonates have been detected at Nili Fossae (Ehlmann et al., 2008b). Komatiite lavas have previously been proposed as possible analogs for Martian rocks on geochemical (Baird and Clark, 1981), morphological (Reyes and Christensen, 1994) and spectral (Mustard et al., 1993) grounds. The komatiite layer detected in the North Pole Dome was associated with talc-carbonate alteration which has been hypothesized to be the result of hydrothermal alteration (Brown et al., 2005).

The North Pole Dome also displays some of the oldest evidence of life in the form of stromatolites (Walter et al., 1980) and microfossils (Ueno et al., 2004) that have been likely to have formed in a volcanic plateau setting where abundant hydrothermal activity is in evidence (Van Kranendonk et al., 2008).

## Carbonates on Mars

Carbonates ($X$-$CO_3$ minerals) are expected to form from basalt in an aqueous alteration environment under a $CO_2$ rich atmosphere (O'Connor, 1968; Booth and Kieffer, 1978; Gooding, 1978; Catling, 1999; Morse and Marion, 1999; Longhi and Takahashi, 2006; Quinn et al., 2006). Because of this, they have been searched for extensively on Mars using remote sensing methods (McKay and Nedell, 1988; Blaney and McCord, 1989; Pollack et al., 1990; Bell et al., 1994; Calvin et al., 1994; Lellouch et al., 2000; Jouglet et al., 2007). Until recently, only trace amounts of carbonate have been detected in bright Martian dust (Wagner and Schade, 1996; Bandfield et al., 2003; Boynton et al., 2009; Palomba et al., 2009) and in Martian meteorites (Bishop et al., 1998a; Bishop et al., 1998b; Bridges et al., 2001; Niles et al., 2009). Recent findings of carbonate by CRISM on MRO (Ehlmann et al., 2008b) and the rover Spirit (Morris et al., 2010) have been the strongest evidence yet for carbonates on Mars.

# METHODS

## Visible to Near Infrared Spectroscopy

CRISM is a visible and infrared imaging hyperspectral spectrometer covering the 0.36-3.92 µm region with 6.55 nm/channel resolution (Murchie et al., 2007). In high resolution targeted mode (relevant to all observations discussed in this paper) CRISM has a ground sampling distance of 15-19m/pixel, and a swath width of approximately 10.8km on the ground.

HyMap is a visible to near infrared (0.4-2.5 µm) imaging spectrometer with 126 spectral bands and a spectral resolution of ~15 nm, manufactured by Integrated Spectronics (www.intspec.com). The spatial resolution of the Pilbara dataset was 5m (Brown et al., 2006). Ground-truth spectra were also taken by the PIMA SP (first generation) handheld spectrometer manufactured by Integrated Spectronics (spectral coverage 1.3-2.5

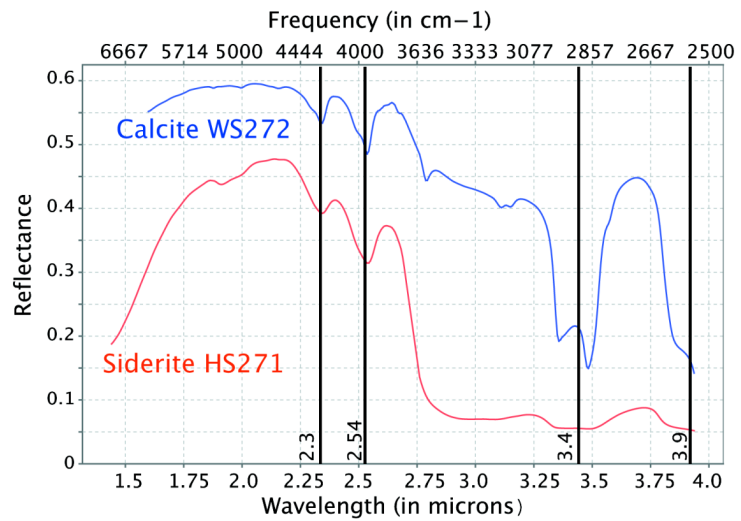

Figure 2. USGS Spectral Library (splib06 from http://speclab.cr.usgs.gov/) near infrared spectra of calcite and siderite showing bands at 2.3, 2.54, 3.4 and 3.9 µms (in addition to other bands). Note apparent weakness of 3.4 and 3.9 µm band in siderite. The average grain size for the calcite sample is 410 microns, and the siderite sample is the 74-250 micron sieved fraction.

µm, 600 bands and resolution of 2 nm).

CRISM and the Observatoire pour la Mineralogie, l'Eau, les Glaces, et l'Activite (OMEGA) were designed to cover the spectral regions where several carbonate absorption bands are located. Figure 2 shows library spectra for several carbonate minerals (resampled to CRISM wavelengths) showing absorption bands at 2.3-2.35, 2.50-2.55, a doublet at 3.3-3.55 and another doublet at 3.8-4.0 µm (Clark et al., 2007).

Clark et al. (1990) assign the 2.3-2.35 µm band to a triple overtone of the asymmetric stretch of the $CO_3^{2-}$ carbonate ion ($3\nu_3$) and the 2.5-2.55 µm band to a combination of the symmetric and asymmetric stretch ($\nu_1+2\nu_3$). We infer here that the 3.3-3.5 µm doublet is due to the second overtone of the asymmetric stretch ($2\nu_3$) and according to Clark et al. the 3.8-4.0 µm doublet is also due to a combination of the symmetric and asymmetric stretch ($\nu_1+\nu_3$). The library spectrum of calcite ($CaCO_3$) demonstrates relatively strong absorptions at all four bands (Figure 2). However, in siderite ($FeCO_3$) the 3.3-3.55 and 3.8-4.0 µm doublet bands are diminished due to low overall reflectivity at wavelengths greater than ~3 µm. This behavior is also true for dolomite and Mg-carbonate. In addition, due to the different sizes of the cations, and resultant changes in electron orbital (bond) lengths (and therefore vibrational frequencies), the carbonate type can be determined by the central position of the 2.3 and 2.5 µm bands. The band positions characteristic of Ca, Fe and Mg-bearing carbonates are centered at 2.34 and 2.54 (Ca), 2.33 and 2.53 (Fe), 2.30 and 2.50 µm (Mg) (Hunt and Salisbury, 1971; White, 1974; Gaffey, 1987).

**Previous Carbonate detections in the Nili Fossae Region**

Mg-carbonate has been inferred in Nili Fossae by recognition of bands at 2.30 and 2.50 µm, accompanied by weak 3.4 and 3.9 µm bands (Ehlmann et al., 2008b). A 1.9 µm $H_2O$ absorption band is also present, probably due to a second, hydrated mineral (e.g., hydrous carbonate (Ehlmann et al., 2008b) or phyllosilicate).

In the low albedo, volcanic region of Syrtis Major, and in particular Nili Fossae, typical maximum visible reflectance values vary from 0.1-0.2. Figure 3 shows example spectra showing 2.3 and 2.54 µm bands from CRISM image HRL 40FF over Jezero Crater. These spectra have been atmospherically 'corrected' using the volcano-scan method (McGuire et al., 2009). The CRISM known noisy band at 2700nm has been removed and 5 x 5 regions have been averaged to obtain the spectra.

We now look at a further parameter that can be used to distinguish carbonate minerals that tends to confirm the recognition of Mg-carbonate – the symmetry of the absorption bands, and combine this information with that of the central absorption band position.

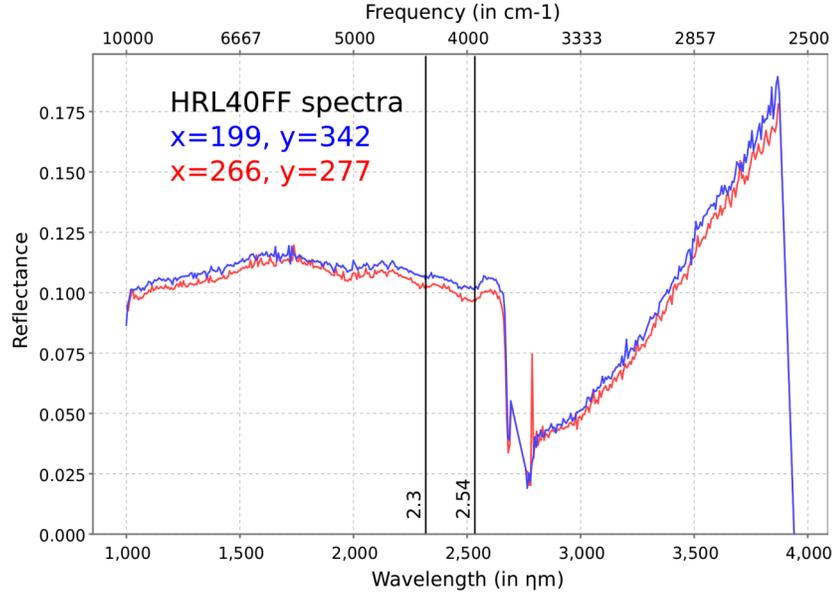

Figure 3 – Example visible to near infrared (L-Channel) CRISM spectra from image HRL 40FF in Nili Fossae. Carbonate 2.3 and 2.54 μm bands are evident. Spectra were corrected for solar incidence angle and atmospheric bands were removed using the standard CRISM 'volcano scan' method (McGuire et al., 2009). The known noisy CRISM spectral channel closest to 2700nm was omitted for clarity and 5x5 averages around the indicated points were taken.

## 2.3 and 2.5 μm absorption band fitting

We have carried out a Nelder and Mead (1965) (also known as "Amoeba") simplex routine to estimate the best fits of an asymmetric Gaussian shape to each of the 2.3 and 2.5 μm bands in the CRISM ratioed spectra. We chose 'shoulders' at 2.2 and 2.384 μm for the 2.3 μm band and at 2.403 and 2.588 μm for the 2.5 μm band. A straight-line continuum was removed from the between the 'shoulders' and the spectrum was then converted to apparent absorbance (Brown, 2006a; Brown et al., 2006; Brown et al., 2008b), before fitting with an asymmetric Gaussian shape. To eliminate unwanted noise the CRISM spectra were treated with a Savitzky-Golay smoothing algorithm before processing (Savitzky and Golay, 1964). An example of the fit is shown in Figure 4. The asymmetric Gaussian shape can be described by the function:

$$\text{If } \lambda \le \lambda_0, \quad f(\lambda) = \alpha \exp\left(-\left[\frac{\lambda - \lambda_0}{\sigma^2}\right]^2\right) \qquad \text{if } \lambda > \lambda_0, \quad f(\lambda) = \alpha \exp\left(-\left[\frac{\lambda - \lambda_0}{(\chi\sigma)^2}\right]^2\right) \qquad (1)$$

We report here the centroid $\lambda_0$, amplitude $\alpha$, half width half maximum (HWHM in μm) $\sigma$, and the asymmetry parameter $\chi$, all to three decimal places (or 1 nanometer) accuracy. Since CRISM bands are spaced roughly 6.6nm apart, 1 nm represents 15% of the distance between spectral bands. The asymmetry parameter is unbounded – values less than 1 indicate 'right asymmetry' (the HWHM on the right side is less than the HWHM on the left side of the centroid, as seen in Figure 4), values greater than 1 indicate left asymmetry.

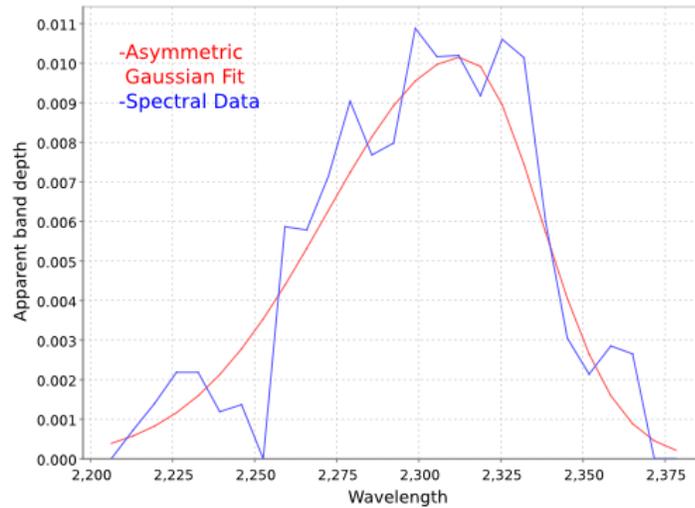

Figure 4. Asymmetric Gaussian fit to the 2.3 μm band in HRL40FF at x=266, y=275 (the parameters for this fit are in the second line of Table 1). The plot is 'upside down' relative to Figure 3 because it is expressed in positive 'apparent band depth' during the algorithmic processing.

We have carried out the same procedure on spectra of candidate minerals from the USGS Spectral Library v6 (Clark et al., 2007), UWinnipeg's Spectroscopy lab (http://psf.uwinnipeg.edu.ca) and a magnesite sample spectra from Janice Bishop's library (Perry et al., 2010). Where possible we chose the purest samples as described in the accompanying documentation. We have also carried out the asymmetric Gaussian fitting on a PIMA spectrum of a stromatolitic carbonate sample from the 'Trendall' locality in the Strelley Pool Chert unit (Brown et al., 2005). Only the 2.3 μm band is analysed for this sample because unfortunately the PIMA's spectral coverage ends at 2.5 μm so we were unable to run the analysis on the 2.5 μm band, which is cutoff. The results of the fitting are shown in Table 1 and the ranges are plotted in Figure 5.

| Spectrum | Source | 2.3band (shoulders 2.2/2.384) | | | | 2.5band (shoulders 2.403/2.588) | | | |
|---|---|---|---|---|---|---|---|---|---|
| | | Cent. | Height | HWHM | Asymm | Cent. | Height | HWHM | Asymm |
| 40FF x=199,y=342 | CRISM | 2.315 | 0.096 | 0.069 | 0.516 | 2.528 | 0.143 | 0.083 | 0.481 |
| 40FF x=266,y=275 | CRISM | 2.312 | 0.100 | 0.059 | 0.531 | 2.521 | 0.147 | 0.085 | 0.480 |
| Pilbara Trendall carbonate by PIMA | Field spectrum (A.Brown) | 2.320 | 0.272 | 0.061 | 0.416 | - | - | - | - |
| ~Library Spectra~ | | | | | | | | | |
| Hydromagnesite CRB208 | UWinnipeg | 2.303 | 0.179 | 0.053 | 0.471 | 2.501 | 0.277 | 0.065 | 0.466 |
| Huntite CRB115 | UWinnipeg | 2.345 | 0.079 | 0.119 | 0.091 | 2.503 | 0.153 | 0.068 | 0.558 |
| Magnesite JB946E | RELAB (J. Bishop) | 2.300 | 0.273 | 0.065 | 0.458 | 2.504 | 0.348 | 0.070 | 0.440 |
| Ankerite 897s285a | RELAB (J. Bishop) | 2.345 | 0.173 | 0.053 | 0.351 | 2.541 | 0.232 | 0.068 | 0.311 |
| Siderite HS271.3B | USGS | 2.337 | 0.209 | 0.070 | 0.516 | 2.537 | 0.308 | 0.071 | 0.557 |
| Dolomite HS102.3B | USGS | 2.325 | 0.309 | 0.067 | 0.292 | 2.519 | 0.427 | 0.074 | 0.345 |
| Calcite WS272 | USGS | 2.346 | 0.184 | 0.070 | 0.256 | 2.541 | 0.237 | 0.064 | 0.238 |
| Analcime GDS1 | USGS | 2.135 | 0.117 | 0.031 | 0.949 | 2.511 | 0.513 | 0.069 | 0.967 |
| Talc HS21.3B * | USGS | 2.313 | 0.400 | 0.027 | 0.510 | 2.467 | 0.180 | 0.012 | 0.676 |
| Brucite HS247.3B | USGS | 2.325 | 0.489 | 0.123 | 0.177 | 2.463 | 0.678 | 0.045 | 1.573 |
| Serpentine HS8.3 ** | USGS | 2.333 | 0.331 | 0.058 | 0.254 | 2.521 | 0.261 | 0.121 | 0.579 |
| Nontronite NG-1 † | USGS | 2.287 | 0.354 | 0.018 | 1.055 | 2.497 | 0.134 | 0.017 | 2.400 |

Table 1 – Results of asymmetric curve fitting. See text for details of processing.
* Note Talc also has a deep absorption band at 2.393(0.264/0.021/0.389) that does not appear in the CRISM 40FF spectra and is not a good match.
** The 2.5 band in serpentine appears to be made up of 3 narrow bands rather than 1 broad band.
† Nontronite has a band at 2.401 (0.181/0.024/0.648) that does not appear in the 40FF spectra.

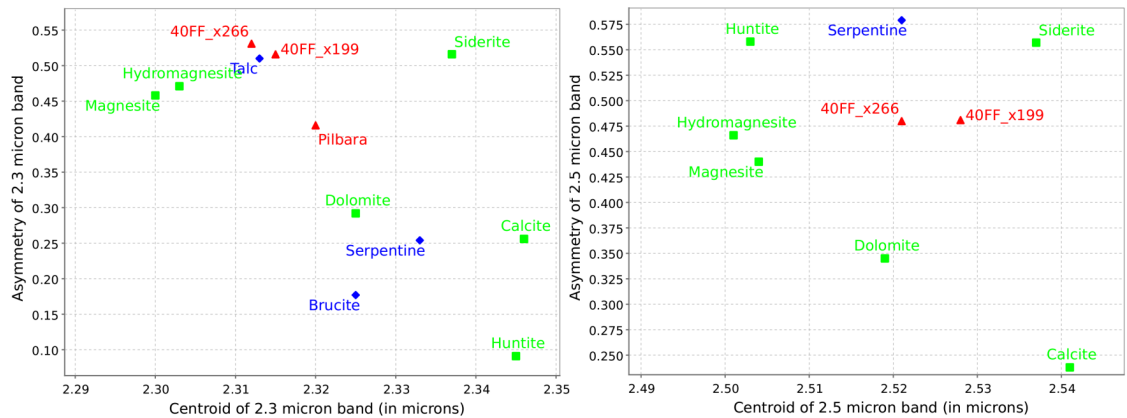

Figure 5. (a) Plot of Centroid vs. Asymmetry for 2.3 bands (b) Plot of Centroid vs. Asymmetry for 2.5 μm bands. Mars and Pilbara spectra are in red, carbonates are in green, other minerals are in blue.

Several ideas are suggested by the comparison of absorption band positions and asymmetry (Figure 5). The 2.3 μm band of talc is a close fit to the CRISM spectra, however it does not have a 2.5 μm band. The Pilbara carbonate sample is the best fit to the band position of the 2.3 μm band. Dolomite, $CaMg(CO_3)_2$ is a reasonable fit to the 2.3 μm band and has a very similar centroid position of the 2.5 μm band. Hydromagnesite ($Mg_5(CO_3)_4(OH)_2.4H_2O$) and magnesite ($MgCO_3$) are also reasonable fits to the shape and position of the 2.3 and 2.5 μm bands, although their centroid positions are consistently low, strongly suggesting that the Martian carbonate detected by CRISM is not due to pure Mg-bearing carbonates. It seems possible that both the 2.3 and 2.5 μm bands may be caused by dolomite, or some form of solid solution of magnesite with siderite, $FeCO_3$.

Other hydrous carbonate minerals that may be of interest (but which we do not have spectral data for) include $MgCO_3.3H_2O$ (nesquehonite) and $MgCO_3.5H_2O$ (lansfordite).

## RESULTS

### Absorption band maps

In order to demonstrate the power of the absorption band mapping technique we have applied it to several CRISM scenes of Nili Fossae containing Mg-carbonate spectra. We used a threshold method to map the presence of carbonates. We required that 1.) the 2.3 and 2.5 μm bands be present, 2.) the 2.3 μm band had a depth of 0.1 and lay between 2.285 and 2.335 and 3.) the 2.5 μm band had a depth of 0.1 and lay between 2.46 and 2.54 μm. The 0.1 threshold depth heuristic was chosen in order to avoid false detections on noisy pixels, and was established by experimentation with a set of CRISM images and hand-checking selected spectra for the presence of recognizable carbonate bands at 2.3 and 2.5 microns. In addition, a simple noise detection algorithm was used before band fitting to eliminate spectra that had obvious noise spikes before processing.

The spectral smile of the CRISM instrument was compensated for by using the smile-corrected two-dimensional CDR WA wavelength tables supplied by the CRISM team. No atmospheric correction was applied before absorption band mapping, since no atmospheric correction is perfect and in this way will avoid any questions over bias introduced by a particular atmospheric removal process. No known gaseous absorption bands are likely to contaminate the 2.3 and 2.5 bands. Two factors help eliminate the misidentification of carbonates due to gaseous absorption bands – 1.) If a well-mixed gaseous absorption band was present then it would be expected to be present throughout an image, not in patches of the image, and 2.) no gas expected in the Martian atmosphere has absorption bands at 2.3 and 2.5 microns of roughly equal strength, which is what the absorption band algorithm is tuned to find.

We applied the spectral absorption band algorithm to all CRISM full resolution (FRT) and half resolution (HRL) images in the eastern Nili Fossae Region, a box from 72.5-85°E and 17-24°N. The images covered a time span from close to the start of the MRO observing phase, October 3, 2006 to July 29, 2009 (the end of this period coincided with the start of an extended MRO idle period due to mission operations fault finding). A total of 72 FRT and 19 HRL images were analyzed. Of these, we assessed 10 FRT and 1 HRL displayed convincing 2.3 and 2.5 µm carbonate bands.

From the images displaying carbonate signatures, we have chosen a subset based on the number of carbonate pixels identified. We show the maps of the 2.3 µm absorption band strength in Figure 6. The 2.3 µm absorption band strength (amplitude) is a proxy for 'apparent carbonate abundance'. We use the term 'apparent abundance' because the strength of an absorption band can be affected by grain size and is strongly affected by mixing with other materials, such as dust.

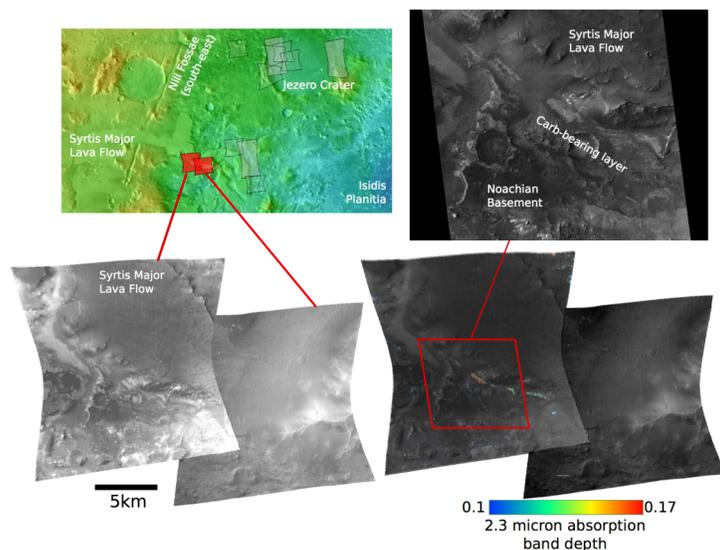

Figure 6 – Absorption band maps from CRISM scenes in Nili Fossae showing depth of the 2.3 µm absorption band. (top left) THEMIS+MOLA context map showing location of images, Jezero Crater and Isidis Planitia. (bottom left) FRT CBE5 (left) and FRT BDA8 (right) using 1.3 µm band. (bottom right) 2.3 µm absorption band for carbonate detections overlain on 1.3 µm band image. (top right) HiRISE image PSP_010206_1976 taken at the same time as FRT CBE5.

Following on from earlier studies (Ehlmann et al., 2008b) we have selected more recent images to examine that show carbonate in a similar setting to the talc-carbonate layers in the Pilbara. FRT BDA8 and FRT CBE5 are located in south eastern Nili Fossae region (Figure 6). The absorption band map highlights a carbonate-bearing layer that is stratigraphically beneath the flat, smooth Syrtis Major volcanic flow units. As in the Pilbara, the talc carbonate layer is sub-parallel to the volcanic flow, and has most likely become exhumed after erosion of the Syrtis Major volcanic flow unit. As in the Pilbara, the hydrothermal talc-carbonate alteration heat source was likely a later volcanic flow unit (if it was not syngentic). The question of the ultimate source of the talc-carbonate alteration is an important question that can only realistically be tackled using data from landed instruments.

## DISCUSSION A: Spectral Interpretations

### Talc vs. Saponite (Mg-phyllosilicate)

In addition to the Mg-carbonate found at Nili Fossae, Elhmann et al. (2009) has discovered a stratigraphically lower Mg-phyllosilicate-bearing unit in many locations throughout Nili Fossae. The Mg-phyllosilicate has been associated with the mineral saponite, a plausible spectral match to the 2.315 and weak 2.39 µm bands (Figure 7). On the basis of experience working with clay-carbonate alteration in the Pilbara region of Western Australia, we have found from HyMap airborne observations that a prominent 2.31 and 2.39 µm are most often associated with talc. This mineral is easily detectable from airborne observations, even when mixed with other minerals (Brown et al., 2004a; Brown et al., 2005) For example, talc often hides carbonate signatures, see also (Dalton et al., 2004). Example HyMap and PIMA spectra from the Pilbara talc deposits, Nili Fossae Mars (FRT 28BA) and USGS library spectra of talc and saponite are presented in Figure 7. Due to the very close accordance of the absorption bands of talc and saponite at the resolution of the CRISM observations, it is not possible at this time in our judgment to discriminate between the two Mg-phyllosilicates saponite and talc. However, on the basis of Earth analog field observations, we favor talc, particularly as a typical alteration product of olivine rich rock units.

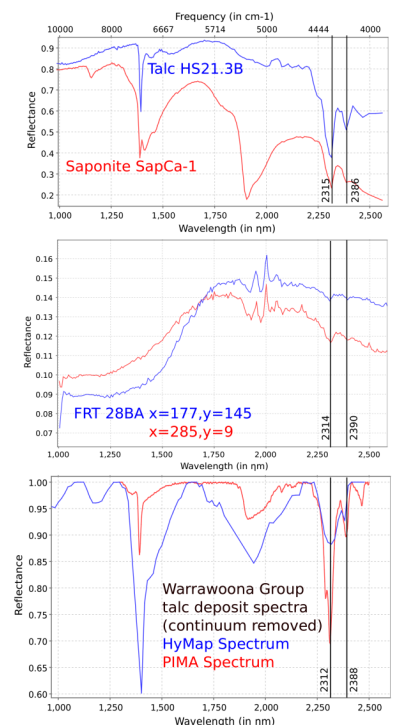

Figure 7 – (top) USGS library spectra of talc and saponite. (middle) Example spectra from FRT 28BA in Nili Fossae after 'volcano scan' atmospheric correction. The spectra are 5x5 averages centered at the x,y locations (in instrument coordinates) shown. Note bands at ~2.315 and ~2.390 µm. The small differences in band center can be explained by ~0.007 µm gap between CRISM bands at full spectral resolution. (bottom) Continuum removed PIMA and HyMap spectrum of talc deposit in Warrawoona Group, Pilbara, Western Australia.

**Other candidate minerals to explain 2.3-2.5 µm bands**

Several other candidate minerals might explain this combination of absorption bands, including other zeolites such as analcime ($NaAlSi_2O_6 \bullet H_2O$), which does not match the 2.3 and 2.5 µm bands quite as well (analcime has a very weak band at 2.13 µm rather than a strong band at 2.3 µm) as a carbonate mineral (Table 1). Another distinct possibility is that two (or more) minerals are contributing to these distinctive 2.3 and 2.5 µm signatures, for example, talc and zeolite mixed together may create the 2.3 and 2.5 µm bands. In our experience, particularly in the Pilbara Project, the only way to resolve this identification is to obtain in-situ ground-truth data from a landed mission with XRD capability such as MSL (Sarrazin et al., 2005).

In this paper we have implicitly assumed the 2.3 and 2.5 µm bands are due to one mineral because the bands appear to occur at similar strengths and do not vary in strength spatially like a mixture of minerals creating these bands would be expected to do. Nevertheless, although we support the hypothesis that carbonate minerals are causing these absorption bands, the possibility exists that the bands may be caused by a mixture of non-carbonate minerals (e.g. talc and zeolite, which has a 2.5 µm band). Future laboratory work at Martian pressure and temperature conditions on metastable hydrated sulfates or zeolites may help shed light on this issue.

# DISCUSSION B: GEOLOGIC IMPLICATIONS

**Geological scenario for carbonates in Nili Fossae**

The carbonate-bearing units of Nili Fossae are relatively reflective in the visible part of the spectrum, commonly fractured, and are capped by an unaltered mafic cap unit (Ehlmann et al., 2008b). Below the carbonate-bearing unit lies a regionally extensive, Fe-Mg smectite-bearing stratigraphic rock unit (Mangold et al., 2007; Mustard et al., 2007). This Fe-Mg smectite unit is the unit we associate with talc, at least in the eastern Nili Fossae region where we have investigated thus far. In the western Nili Fossae region, the Fe-Mg smectite unit cannot be composed of talc entirely because there are regions where only Fe-smectite is found (with an absorption band centered at 2.29 micron (Mangold et al., 2007)). Above the carbonate unit, an Al-phyllosilicate unit is sometimes observed, always stratigraphically higher than the carbonate unit. Ehlmann et al. (2008b) pointed out that the carbonate unit also appears stratigraphically coincident with olivine-bearing units observed by Mustard et al. (2008).

Talc-carbonate propylitic zone alteration reactions. A well-known two-step representative alteration reaction leading to a propylitic style of mineral assemblage is given in Table 2. This reaction is often associated with the outermost, lower temperature (<150ºC) propylitic zones in hydrothermal systems (Lowell and Rona, 1985). In many locations (e.g. CRISM spectra in Figure 7 from FRT 28BA) a deep and wide 1 µm band is present in the pixel with talc or carbonate bands. This suggests olivine is co-located with the talc-carbonate layers. This could be explained by: 1.) olivine sands overlying the talc or carbonate layers, or 2.) olivine may be preserved within the talc or carbonate layers. If scenario 2.) is correct, the presence of olivine and its alteration product within the same layers would suggest that not enough heat and water was available for the alteration reaction to run to completion. This contrasts with the situation in the 3.5 Ga Warrawoona group mafic/ultramafics, where olivine signatures are no longer visible from airborne observations and olivine phenocrysts have usually been totally replaced when viewed in thin section (Brown, 2006b).

| Basalt Clay-Carbonate Two-step Alteration Reaction | | |
|---|---|---|
| Mg-rich precursor | *Step 1* | $3\,Mg_2SiO_4 + SiO_2 + 2H_2O \Rightarrow 2Mg_3Si_2O_5(OH)_4$ <br> *forsterite     aqueous silica      serpentine* |
| | *Alt. Step 1a* | $18Mg_2SiO_4 + 6\,Fe_2SiO_4 + CO_2 + 26H_2O \Rightarrow 2Mg_3Si_2O_5(OH)_4 + 12Fe_3O_4 + CH_4$ <br> *forsterite     fayalite                          serpentine     magnetite* |
| | *Step 2* | $2Mg_3Si_2O_5(OH)_4 + 3CO_2 \Rightarrow Mg_3Si_4O_{10}(OH)_2 + H_2O + 3MgCO_3$ <br> *serpentine                   talc                  magnesite* |
| Fe-rich precursor | | $Fe_2SiO_4 + 2H_2CO_3 \Rightarrow \quad SiO_2 + 2FeCO_3 + 2H_2O$ <br> *fayalite  carbonic acid  silica    siderite* |
| Alternative one step reaction | | |
| High Mg conditions with $CO_2$ | | $Fe_2SiO_4 + Mg_2SiO_4 + nH_2O + CO_2 \rightarrow Mg_3Si_2O_5(OH)_4 + Fe_3O_4 + MgCO_3 + SiO_2$ <br> *fayalite  forsterite                          serpentine           magnetite  magnesite* |

Table 2 – Two step clay-carbonate reaction series to arrive at carbonates from an Mg-rich ultramafic precursor. Fe-rich precursors (from fayalite) may also be applicable on Mars, although siderite has not yet been detected by CRISM. Alternate Step 1a shows it is possible in the presence of $CO_2$ to form methane as part of the serpentinization process. No attempt has been made to balance the one step reaction.

The formation of carbonate such as magnesite is enhanced by high dissolved $CO_2$ (>10%) in the metasomatic fluids and higher magnesian ultramafics. High magnesian basalt is likely to facilitate alteration fluids that are neutral to basic (Moody, 1976; Donaldson, 1981; Welhan, 1988). This suggests alteration by circum-neutral, low temperature (<150ºC) and short-lived hydrothermal conditions would be in accord with the current state of knowledge of the Nili Fossae carbonate unit.

**Sedimentary vs. hydrothermal modes of formation**

Catling (1999) suggested a model for carbonate formation in an evaporation sequence under a high partial pressure of $CO_2$. It has repeatedly been suggested that a putative Martian ocean in an atmosphere with a high partial pressure of $CO_2$ would have favored the creation of large volumes of sedimentary carbonate deposits (Morse and Marion, 1999; Longhi and Takahashi, 2006).

A hydrothermal alteration mode of formation for the carbonates in Nili Fossae would not have required a long standing ocean (or even lakes), and could have occurred in a spatially restricted region. Unlike the Catling and Morse-Marion ocean scenarios, the hypothesis we have presented here would not require the presence of large oceans in Noachian Mars or a warmer atmospheric environment.

Hydrothermal formation of carbonates on Mars has been suggested previously (Griffith and Shock, 1995; Treiman et al., 2002). By analogy with the Earth, carbonates associated with hydrothermal activity would be ideal locations for the creation and preservation of microfossils (Walter and Des Marais, 1993; Farmer and Des Marais, 1999), therefore the Nili Fossae clay-carbonate sequence may be a good location to search for biomarkers in future landed Astrobiology missions to Mars.

**Further implications – dust, warmer early Mars, acid alteration, and methane**

<u>Nili Fossae as a possible 'carbonate dust' source</u>. As pointed out by Ehlmann (2008a), carbonate at Nili Fossae may be a source region for the carbonate found in the ubiquitous Martian dust by the Thermal Emission Spectrometer (TES) instrument (Bandfield et al., 2003) and its formation process may be linked to the preterrestrial carbonate formed within ALH84001 (Romanek et al., 1994; McKay et al., 1996), Nakhla (Gooding et al., 1991) and Chassigny (Wentworth and Gooding, 1991; Wentworth and Gooding, 1994).

<u>Not enough carbonate yet to support a warmer, wetter Early Mars</u>. A large carbonate reservoir on Mars has long been sought to support the hypothesis of a warmer, wetter early Mars (Kahn, 1985; Pollack et al., 1987; Schaefer, 1990; Schaefer, 1993; Haberle et al., 1994; Jakosky et al., 1994; Carr, 1999). Crustal carbonates may have acted as a sink for $CO_2$ at a time when the atmosphere was thicker. The small outcrops of carbonate minerals detected thus far by CRISM appear unlikely to constitute a large enough sink for atmospheric $CO_2$, however the possibility still exists that they may be the first detected hint of larger deposits that are not currently exposed.

<u>Lack of acidic alteration at Nili Fossae</u>. It has been suggested that Mars went through periods where water was more abundant (for example, in the Noachian when large amounts of phyllosilicates were emplaced, for example at Mawrth Valles (Bibring et al., 2006; Loizeau et al., 2007; Loizeau et al., 2010)) and following this, a period of acid-sulfate alteration took place (perhaps before or during the Hesperian period) as water was lost from the surface of the planet (Fairen et al., 2004; Moore, 2004). CRISM and OMEGA do not detect large abundances of sulfate minerals in the Nili Fossae region, although many different types of phyllosilicate minerals have been identified (Ehlmann et al., 2007; Ehlmann et al., 2008b; Ehlmann et al., 2010).

The presence of carbonate and relative lack of sulfate at Nili Fossae could be explained by two different scenarios: 1.) the carbonate unit was formed in neutral subsurface conditions and was exhumed after global acid-sulfate alteration abated or 2.) the Nili Fossae region was not exposed to extensive acid-sulfate alteration (Ehlmann et al., 2008b). At this time there is not enough information to hand to decide between these alternatives.

Possible link to methane detection. Recent telescopic observations of Mars have been used to suggest methane may outgas seasonally from the Nili Fossae region (Mumma et al., 2009), although these observations have been called into question on the basis of possible telluric contamination (Zahnle et al., 2010). If indeed the Nili Fossae region is a source of methane, the serpentinization process outlined in Table 2 (Step 1a) is one potential source. Oze and Sharma (2005) have suggested that for serpentinization to occur on Mars it is necessary to have: 1.) pressure of 0.5 bar (i.e. subsurface on Mars), 2.) temperature < 330°C and 3.) liquid water and dissolved $CO_2(aq)$ must be present. As shown in Table 2, this proposed reaction would be consistent with ongoing clay-carbonate hydrothermal alteration, however for this process to be a source of methane on Mars today, it would imply that clay carbonate alteration is occurring kms below the Martian surface, and cannot be directly linked to the Nili Fossae carbonate observed by CRISM. Any linkages between current day methane production and carbonate likely requires further investigation from landed spacecraft.

## CONCLUSIONS

The results of our work are summarized below.

1. We have found that based on the position of the 2.3 and 2.5 micron band the carbonate in Nili Fossae are unlikely to be pure Mg-carbonates, and are likely to be a solid solution with siderite or calcite, and may be partially dolomitised magnesite or hydromagnesite.

2. We propose that a talc-carbonate mineral assemblage similar to that present at the Archean Warrawoona Group in the Pilbara region of Western Australia is a plausible analog for the clay and carbonate-bearing Noachian rock unit at Nili Fossae, Mars.

3. We suggest that hydrothermal alteration of a basaltic mafic/ultramafic precursor has produced the carbonate and clay detected by CRISM in Nili Fossae. This formation scenario requires liquid water, but would not necessitate large lakes or oceans on Noachian Mars.

4. The presence of clay and carbonate in the Nili Fossae region suggests that biomarkers (if present) could have been preserved within these rocks, as they have been in the Pilbara region.

Further investigation of this area by landed missions with mineralogical characterization capability (VNIR, Raman or XRD) would enable a more detailed analysis of the clay and carbonate components of these rock units and should be a high priority of the Mars Exploration Program.

# ACKNOWLEDGEMENTS

We would like to thank Bethany Ehlmann, Ed Cloutis, Katrin Stephan and Nicolas Mangold for their reviews which improved this paper immensely. We would also like to thank the entire CRISM Team, particularly the Science Operations team at JHU APL. This work was supported by the NASA Interdisciplinary Exploration Systems Program NNH05ZDA001N.

# APPENDIX

## 3.4 μm band sensitivity analysis

We present a sensitivity analysis of the 3-4 μm region for CRISM in order to assess absorption band strengths that would be required for band recognition in the 3-4 μm region.

The OMEGA team considered the detection of carbonates to require detection of a 3.4 μm band (Jouglet et al., 2007) and the CRISM carbonate spectra do not show this band. Ehlmann et al. (2008b) suggest that laboratory spectra of carbonates show the 3.4 and 3.9 μm bands of carbonate can be reduced or eliminated in the presence of water, coatings or other minerals. Hydrous carbonates frequently have reduced or eliminated 3.4 or 3.9 μm bands (Cloutis et al., 2000).

Here we determine the relationship between signal to noise and band depth in order to determine the relative sensitivity of the CRISM instrument to the 3.4 and 2.5 μm bands. We chose the calcite spectral library as an exemplar, since it has a strong 3.4 μm band (compared to siderite, for example, as in Figure 2).

From Table 1, the calcite HWHM of the 2.5 μm band is 0.064 μm. We estimate the HWHM of the calcite 3.4 μm band to be 0.130 μm. These figures are somewhat dependant on the purity of samples, grain size effects and viewing conditions. However, we would assert that we cannot envisage a scenario where the carbonate 3.4 μm absorption band is likely to be more than 3 times wider than the 2.5 μm band.

Following Kirkland et al. (2001), we calculated a "relative detection limit" incorporating the signal to noise of the CRISM instrument in the 2.5 and 3.4 µm regions. Using the formula in Equation 2:

$$DL = \frac{100x \text{ confidence factor}}{\frac{signal}{noise_{PtoP}} \Big/ 2 \times \sqrt{\frac{band \text{ } FWHM}{sampling \text{ } interval}}} \qquad (2)$$

and using a sampling interval of 6.6nm, approximate signal to noise (peak to peak or PtoP) for CRISM at 2.5µm of 450 and at 3.4µm of 80 (Murchie et al., 2007) (and eliminating the somewhat arbitrary confidence factor by setting it to 1) we find the 2.5 µm feature has a detection limit of 0.05 and the 3.4 µm feature has a detection limit of 0.199.

Considering only the relative detection limits, this implies that the CRISM spectral band detection threshold for the 3.4 µm feature is four times that of the 2.5 µm feature. In the limit of small abundances (or low albedo conditions due to highly absorbing co-contaminants, the likely scenario for Nili Fossae) the 3.4 µm band will become undetectable by CRISM far more quickly than the 2.5 µm band. This may provide an additional explanation for why strong 3.4 (or 3.9 µm) carbonate absorption bands have not been detected by CRISM or OMEGA.